\documentstyle[prl,aps,epsf,twocolumn]{revtex}

\tighten

\begin{document}
\draft
\twocolumn[\hsize\textwidth\columnwidth\hsize\csname @twocolumnfalse\endcsname

\title{\bf Identification of relaxation and diffusion mechanisms in
amorphous silicon}

\author{G.T. Barkema}

\address{ITP, Utrecht University, P.O. Box 80000, 3508 TA Utrecht, the
Netherlands\protect\cite{barkadd}}

\author{Normand Mousseau}

\address{Computational Physics, Faculty of Applied Physics, TU Delft,\\
Lorentzweg 1, 2628 CJ Delft, The Netherlands,\\
and\\
Department of Physics and Astronomy and CMSS, Ohio University, Athens, OH 45701, USA
\protect\cite{mousadd}}

\date{\today}

\maketitle


\begin{abstract}
The dynamics of amorphous silicon at low temperatures can be
characterized by a sequence of discrete activated events, through which
the topological network is locally reorganized. Using the
activation-relaxation technique, we create more than 8000 events,
providing an extensive database of relaxation and diffusion mechanisms.
The generic properties of these events --size, number of atoms
involved, activation energy, etc.-- are discussed and found to be
compatible with experimental data. We introduce a complete and unique
classification of defects based on their topological properties and
apply it to study of events involving only four-fold coordinated atoms.
For these events, we identify and present in detail three dominant
mechanisms.
\end{abstract}

\vskip2pc
]

\vspace*{-0.5cm} \narrowtext

At low temperatures, the microscopic dynamics of materials typically
takes place by activated processes, in which the energy barriers
crossed are high compared to the average thermal energy. In
crystals, the microscopic mechanisms at the origin of the dynamics can
often be found based on symmetry arguments. Disordered materials,
however, require direct simulations in order to identify these
mechanisms.  The standard approach, introduced by Weber and
Stillinger~\cite{weber85}, is to simulate a material at a high
temperature and to quench configurations at regular intervals. If this
temperature is sufficiently high and the time between quenches large
enough, these configurations will be distinct, and their sequence
may provide a basis for the reconstruction of activated dynamics.
Difficulties arise, however, since (1) it is never certain that the
path reproduced is one that the configuration would follow if
simulated at lower temperature and (2) the Meyer-Neldel compensation
rule, affecting the prefactor of the diffusion constant, often favors
different mechanisms at high and low temperatures\cite{boisvert95}.

Using the recently proposed activation-relaxation technique
\cite{art}, it is possible to follow the reaction path as
configurations move from one low-energy structure to another, allowing
barriers on all energy scales to be crossed. We apply this method to
generate a database of structural changes in amorphous silicon that
can then be classified according to their topology, which is
well-defined since the first and second-neighbor shells are completely
separated in high-quality simulation cells.  Besides being an important
material in its own right because of its technological importance,
amorphous silicon is also the archetypical example of the class
of tetrahedrally coordinated continuous-random networks.

Self-diffusion in {\it c}-Si has an activation energy of about $4.5
\pm 0.5$ eV~\cite{frank84}.  For {\it a}-Si, experimentalists find a
range of activation energies indicating a multiplicity of
mechanisms. In differential calorimetric and conductivity measurements
on samples freshly prepared by ion-bombardment, the
average activation barrier is found to vary from 0.23 to 2.7 eV as a
function of relaxation or temperature \cite{roorda91,shin93}.  Since
samples are far from equilibrium, these barriers have to be taken as
lower bounds: barriers should be higher in well relaxed samples,
reaching values similar to that found in {\it c}-Si.

The nature of relaxation and diffusion mechanisms itself is still
controversial: calorimetric and Rahman measurements indicate that
relaxation should be local, involving relaxation of point defects
\cite{roorda91}, while M\"ossbauer experiments suggest that up to
$10^4$ atoms are involved, at least to some degree, in the relaxation
of a single dangling bond~\cite{muller94}.

Theoretical studies of the dynamics of {\it a}-Si have concentrated so far
on very small lattices and have produced only a handful of events. It is
therefore difficult to draw solid conclusions from these works, either
regarding the activation energies, or the mechanisms involved.

In this work, we use the activation-relaxation technique, which can
efficiently produce a large number of events. ART is described in
detail elsewhere \cite{artb}.
In order to produce good structural amorphous silicon, we
proceed as in \cite{art,gaas}, and use the Stillinger-Weber potential
\cite{sw} with a three-body interaction increased by 50\%.  After each
event, the volume of the cubic cell is optimized to minimize the total
energy of the network.  For each of three independent runs, we start
from a different randomly-packed 1000-atom cell relaxed to a local
minimum, and apply ART steps iteratively until the configuration
reaches a good structural quality\cite{art,gaas}.  ART moves that
converge to a saddle-point within at most 600 iterations are either
accepted or rejected, using a Metropolis accept-reject criterion with a
fictitious temperature of 0.25 eV.  The optimization phase typically
requires a couple of events per atom. Just to be sure, we reject the
first 4000 ART moves. After optimization, we make another 6000 ART moves,
ending with a combined database of 8106 events that are examined here.

\begin{figure}
\epsfxsize=8cm
\epsfbox{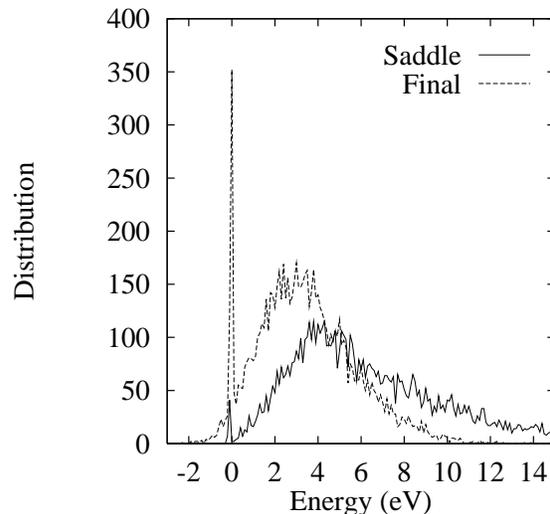}
\caption{
Distribution of energy barrier $B$ (solid line) and energy difference
$\Delta E$ (dotted line) for the 8106 events mentioned in the text.
\label{fig:energy}}
\end{figure}

The distribution of energy barriers $B$, as well as the energy
difference between the initial and final configurations $\Delta
E=E_f-E_i$ is given in Figure \ref{fig:energy} for the whole set of
events.  The precision in the energy barrier is about $\pm 0.5$ eV due
to the approximations used in finding the saddle point \cite{artb}.
The exact shape of these two distributions is a convolution of the
topography of the energy landscape with the biases of the
activation-relaxation technique \cite{artb}; since these biases are not
known, one cannot infer quantitative information from the frequency of
occurrence of certain events, although we might expect a certain
correlation between this and the real frequencies of specific events.

As can be seen in Fig. \ref{fig:energy}, the distribution for the
energy barrier peaks around 4 eV and tails off beyond 15 eV. Although
the high-energy tail of the distribution is unlikely to be sampled in
reality -- its time scale goes much beyond what can be reached
experimentally -- it demonstrates the ease with which ART can explore
the energy landscape of disordered materials.  The continuous spectrum
of energy barriers is in line with experimental data \cite{roorda91}.
Looking more closely, the small peak around 0 eV in the distribution of
the energy barrier is associated with unstable configurations, and the
large peak in the distribution of the energy difference corresponds to
atomic exchanges as discussed below.  As expected for well relaxed
configurations, most activated mechanisms lead to states of higher
energy; only a small proportion of events should lower the energy.

The size of an event, as defined by the number of atoms that move more
than a certain threshold distance, can also provide some insight into
the nature of relaxation.  With a threshold of 0.1 \AA, we find that
the number of atoms displaced from a minimum to a saddle point varies
between 1 and 80, with a peak at about 25. From the initial to the final
minimum, this number runs to above 100 with a maximum
at 40. This result sets the minimal system size for the study of these
mechanisms to several hundreds of atoms to limit boundary effects,
an order of magnitude larger than the typical size of the relaxation
mechanism.  We find that the distribution of the number of atoms
involved is almost independent of the specific topological class studied.

Looking at the topology of the structural changes in more detail, we
concentrate on two classes of events: {\it perfect} events, that
directly involve\footnote{By directly involved, we mean atoms that see
their neighbor list change between the initial and final state.} only
fourfold coordinated atoms, and {\it conserved} events, that do
involve only the diffusion of coordination defects, while the total number
of defects is preserved from the initial to the final state.  There
are 802 perfect events, and another 1979 conserved events; the
remaining 5325 describe the creation and/or annihilation of
coordination defects.

We now consider the 802 perfect events. Table \ref{table} shows first
that contrary to previous conclusions\cite{previous}, stretched bonds
do not play a major role in the relaxation and diffusion mechanisms of
well-relaxed structures. This is in agreement with the narrow spread
of nearest-neighbor distances seen experimentally; long bonds are most
probably an artifact of incompletely relaxed computer-prepared
structures. Table \ref{table} also indicates that the bond-breaking
occurs in a complex process which involves in parallel the formation
of new bonding and a considerable local rearrangement. Such a mechanism
allows the total amount of energy needed at any time to be much
smaller than what a naive evaluation, considering the process of
breaking bonds isolated from that of creation, would give.

\begin{figure}
\epsfxsize=8cm
\epsfbox{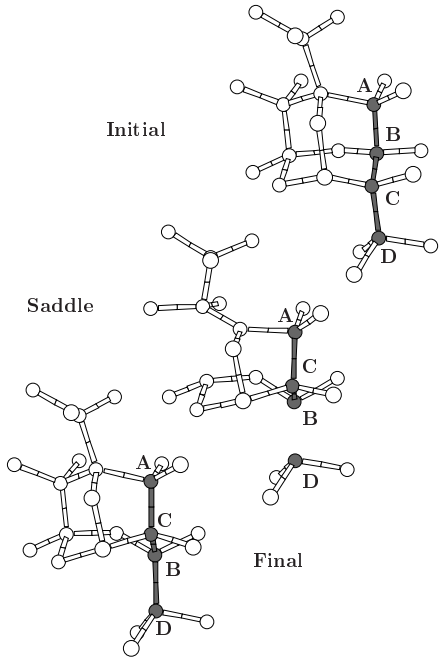}
\caption{Typical example of a bond rearrangement. Only four atoms,
labeled 'A', 'B', 'C', and 'D', change their list of neighbors from
the initial to the final state.
\label{classif}}
\end{figure}

A topological classification provides more insight into the
perfect events. Since the number of broken bonds must be identical to
that created between the initial and final events, and all atoms
changing neighbors must remain fourfold coordinated, the simplest
topological description is to alphabetically label all atoms involved,
and list all bonds between them, present before or after the event, or
both. Figure \ref{classif} shows a typical bond rearrangement.

A more compact classification of the rearrangement is based on the
observation that a loop of alternating old and new bonds exists that
visits all atoms involved (except for the rare cases in which the atoms
involved belong to spatially disconnected groups).  We can generate
such a loop by starting on one atom topologically involved in
the event, and alternatingly following old and new bonds, until the
path returns to the initial atom.  If these atoms are labeled alphabetically,
a loop is characterized by the sequence in which atoms are visited. The
event in figure \ref{classif} could for example be denoted by a loop
``ABCDBC''. Always using the alphabetically lowest notation, in this case
``ABACBD'',  guarantees us a {\it complete and
non-degenerate} classification for all perfect events.  For more
complex events, involving the diffusion or the creation/annihilation of
coordination defects, this algorithm requires the addition of ``ghost
bonds'' between pairs of atoms whose coordination goes up or down,
preserving the completeness and uniqueness of the classification, but
increasing significantly the complexity of the algorithm.

\begin{figure}
\epsfxsize=8cm
\epsfbox{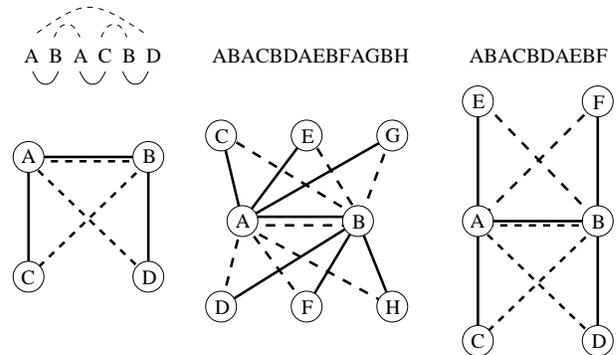}
\caption{
Most frequents events in our simulation. The left figure
corresponds to the WWW bond-switching mechanism, the middle figure
to concerted exchange, the right figure to a double bond-switching
mechanism.
\label{diag1}}
\end{figure}

Applying the classification scheme discussed above to the perfect
events, we find that three mechanisms account for more than 85\% of
the distribution.  The most frequent type of events, 68\% of all, is,
in the notation introduced above, ABACBD and is illustrated in
Figs. \ref{classif} and \ref{diag1}. It corresponds, surprisingly, to
the bond-switching algorithm proposed by Wooten, Winer and Weaire
\cite{www} for the preparation of amorphous silicon from the
crystalline state: two bond-sharing atoms exchange a pair of
neighbors. Peaking a low energy and with a relatively narrow
distribution of activation energy, roughly 2 eV, this event is also
clearly favored in this class.

The second commonest type of perfect event (12\%), in our database, is
classified as ABACBDAEBFAGBH and illustrated in figure \ref{diag1}.
These events involve a perfect switch of two nearest-neighbor atoms.
Such a mechanism, dubbed concerted-exchange (CE), was proposed some
time ago by Pandey to explain self-diffusion in {\it
c}-Si~\cite{pandey86} and its existence is still debated.  The
activation energy distribution for the CE is strongly skewed: the
lowest events happen at 3.6 eV, with a tail that runs up to about
12 eV. The lower part of the distribution is in agreement with the
activation energy calculated for {\it c}-Si using the Stillinger-Weber
potential \cite{kaxiras88} but is higher than for WWW-type 
events.

The CE can be decomposed in three basic WWW bond-switching steps: each
atom exchanges three neighbors with the other atom. As could be
expected, therefore, the third-most frequent mechanism (8\%)
(schematized in Fig. \ref{diag1}) is a double bond-switching mechanism
represented in our classification as ABACBDAEBF. The double WWW
presents an even wider and flatter distribution of activation energies
than the CE mechanism; it starts at about 3.5 eV and runs up to 17 eV.

These three mechanisms account for 703 out of 802 perfect events. All
other mechanisms occurred less than 1\% of the time and involve,
typically, a larger number of bond exchanges, and a higher activation
energy.

Although more than 95 \% of the atoms in our three samples are
four-fold, most events actively involve defects either through
diffusion (conserved events) or creation and/or annihilation.
Topological analysis reveals that no dominant class of event emerges
in either of these cases, due to a much larger number of possible
topological environments that can be found when defects are around.
It is nevertheless still possible to highlight some generic trends.
The creation/annihilation mechanisms and other results from this study
will be presented in a longer paper~\cite{longer}, but we can say a
word about diffusion of defects.

Contrary to conventional experimental imagery, we do not find any
significant presence of crystalline-type defect in these events,
such as interstitials or vacancies. Diffusion is controlled by jumps of
coordination defects that can require a significant reorganization of
the lattice: the number of broken and created bonds is 1.25 bond
higher than in perfect events.  Again, as shown in Table \ref{table},
there is no evidence of diffusion dominated by highly stretched bonds.
One might also expect that defect-mediated relaxation is easier
than fully four-fold coordination. In contrast to this prediction however,
we find that the distribution of activation energies is very 
similar to that of the perfect events, with a peak at
around 4 eV and a tail extending to 14 eV.
Finally, as mentioned previously, the total number of atoms
involved is found to be essentially insensitive to the class of
events.  This might be explained by the fact that more than 60\% of
conserved events have a root based on one of the three mechanisms
dominating the perfect events (Fig. \ref{diag1}).

We have presented here the first detailed analysis of microscopic
diffusion and relaxation mechanisms in {\it a}-Si, introducing a {\it
complete and non-degenerate} topological classification of events.  In
particular, we find that perfect (non-defect-based) diffusion involves
mainly mechanisms corresponding to variations on the bond-switching
procedure of Wooten, Winer and Weaire, which include the
concerted-exchange mechanism proposed by Pandey. The size and
activation energy distributions of events are realistic and provide a
first check with experiment.  We also find that the microscopic
mechanisms responsible for defect-involved diffusion in {\it a}-Si
differ from those in the crystal and are based on jumps of
coordination defects; there is no trace of vacancy- or
interstitial-mediated diffusion. In view of the generic character of
the analysis, these conclusions should apply to other elemental
tetrahedral amorphous semiconductors as well.

All these mechanisms result directly from the activation-relaxation
technique simulations, with no {\it a priori} input provided about the
nature of possible mechanisms.  Much remains to be analyzed from the
large database presented here. In particular, it will be necessary to
re-compute some of these barriers using more accurate interactions
such as tight-binding or LDA functionals, and to analyze in more
detail the nature of defect annihilation/creation.

{\it Acknowledgements.} NM acknowledges partial support by the
Stichting FOM (Fundamenteel Onderzoek der Materie) under the MPR
program.

\bibliographystyle{prsty}

\begin{table}
\caption{Some properties of different mechanisms.  $N$ is the number of
these events in the database; $b_{is}$ and $c_{is}$ refers to the
average number of broken and created bonds between the initial minimum
and the saddle point; $b_{if}$ is the same quantity between initial and
final minimum, and $l^b_{if}$ / $l^c_{if}$ gives their respective
length in \AA; $d$ is the average atomic displacement between initial and
final point (\AA); $B$ and $\Delta E$ correspond to the maxima in the
histograms of the energy barriers and energy changes (eV). The numbers
in brackets are the fluctuations in these quantities. The conserved
events analyzed here are those with an energy barrier smaller than 8 eV.
\label{table} }
\begin{tabular}{l|ccccc}
Type       & Perfect    & WWW       &  CE       & Double    & Conserved \\
\hline
$N$        & 802        & 545       &  94       & 64        & 1147      \\
$b_{is}$   & 3.3 (1.6)  & 2.6 (1.2) & 5.4 (1.3) & 4.3 (1.3) & 3.7 (1.4) \\
$c_{is}$   & 2.8 (1.5)  & 2.2 (1.0) & 4.7 (1.6) & 3.8 (1.5) & 3.4 (1.4) \\
$b_{if}$   & 3.0 (1.6)  & 2.0 (.0)  & 6.0 (.0)  & 4.0 (.0)  & 4.3 (1.6) \\
$l^b_{if}$ & 2.37 (.05) & 2.38 (.05)&2.38 (.03) &2.37 (.04) &2.43 (.09) \\
$l^c_{if}$ & 2.46 (.07) & 2.47 (.07)&2.38 (.03) &2.42 (.05) &2.55 (.10) \\
$d$        &   --       & 2.2 (.25) & 3.3 (.2)  & 3.0 (.2)  & 2.9 (.7) \\
$B$        & 4.1        & 4.0       & 5.8       & 4.0-6.0   & 4.1       \\
$\Delta E$ & 2.4        & 2.2       & 0.0       & 3.0       & 2.3       \\
\end{tabular}
\end{table}

\end{document}